\documentclass[preprint]{aastex}

\usepackage[dvips]{color}
\usepackage{here}
\usepackage{amsmath,amssymb}
\slugcomment{}
\shortauthors{Xue et al.}
\shorttitle{Evolution of the spin-orbit angle}
\begin{document}
\title{Tidal evolution of the spin-orbit angle in exoplanetary systems}
\author{
Yuxin Xue\altaffilmark{1}, 
Yasushi Suto\altaffilmark{1,2,3},
Atsushi Taruya\altaffilmark{1,2,4},
Teruyuki Hirano\altaffilmark{1,5}, \\
Yuka Fujii\altaffilmark{1,6}, and
Kento Masuda\altaffilmark{1}
} 
\altaffiltext{1}{Department of Physics, The University of Tokyo, Tokyo
113-0033, Japan}

\altaffiltext{2}{Research Center for the Early Universe, School of
Science, The University of Tokyo, Tokyo 113-0033, Japan}

\altaffiltext{3}{Princeton University Observatory, Princeton, NJ 08544}

\altaffiltext{4}{Yukawa Institute for Theoretical Physics, Kyoto
University, Kyoto 606-8502, Japan}

\altaffiltext{5}{Department of Earth and Planetary Sciences, Tokyo
Institute of Technology, Tokyo 152-8551, Japan}

\altaffiltext{6}{Earth-Life Science Institute, Tokyo Institute of
Technology, Tokyo 152-8551, Japan}

\email{yuxin@utap.phys.s.u-tokyo.ac.jp}
\begin{abstract}
 The angle between the stellar spin and the planetary orbit axes
(spin-orbit angle) is supposed to carry valuable information on the
initial condition of the planet formation and the subsequent migration
history.  Indeed current observations of the Rossiter-McLaughlin effect
have revealed a wide range of spin-orbit misalignments for transiting
exoplanets. We examine in detail the tidal evolution of a simple system
comprising a Sun-like star and a hot Jupiter adopting the equilibrium
tide and the inertial wave dissipation effects simultaneously.  We find
that the combined tidal model works as a very efficient realignment
mechanism; it predicts three distinct states of the spin-orbit angle
(i.e., parallel, polar, and anti-parallel orbits) for a while, but the
latter two states eventually approach the parallel spin-orbit
configuration. The intermediate spin-orbit angles as measured in recent
observations are difficult to be achieved.  Therefore the current model
cannot reproduce the observed broad distribution of the spin-orbit
angles, at least in its simple form. This indicates that the observed
diversity of the spin-orbit angles may emerge from more complicated
interactions with outer planets and/or may be the consequence of the primordial misalignment between the proto-planetary disk and the stellar spin, which requires future detailed studies.
\end{abstract}

\clearpage

\section{Introduction \label{sec:intro}}

More than 1000 exoplanets discovered so far have exhibited a
surprising diversity in their physical properties. These are important
observational clues to unveil their formation and evolution
processes. In particular, unexpectedly large fractions of hot Jupiters
(giant gaseous planets orbiting the central star with periods less than
1 week) and planets with highly eccentric and/or misaligned (their
orbital axis is not aligned with the spin axis of the central star)
orbits are regarded as serious challenges to conventional models of
planet formation that have been proposed to explain the properties of
our solar system.  According to the standard core-accretion scenario,
such gas giants are supposed to form beyond the ice line in nearly
circular orbits.  Thus it is widely believed that
the discovered hot Jupiters should have formed at a large distance beyond
the ice line, and then somehow migrated towards the orbits close to the
central star \citep[e.g.,][]{Ida2004a,Ida2004b}.

A widely accepted scenario is Type II migration in which gas giants
migrate inward by the planet-disk interaction and halt at the inner edge
of the disk \citep[e.g.,][]{Lin1996}. Other scenarios require dynamical processes after the depletion of the gas disk, including planet-planet scattering
\citep[e.g.,][]{Rasio1996,Nagasawa2008,Nagasawa2011}, secular evolution
\citep{Wu2011}, and the Kozai mechanism \citep{Kozai1962,Wu2003,Fabrycky2007,Naoz2012}. 
While the former normally predicts alignment of the stellar spin axis and the planetary orbital axis, the latter dynamical processes are found to enhance the eccentricity and inclination of planets. Thus, the spin-orbit angle distribution may be a useful discriminator of the planetary migration scenarios. 

The projected angle ($\lambda$) between the stellar spin and the
planetary orbit of transiting planets can be determined through the
Rossiter-McLaughlin (RM) effect \citep{R1924,M1924, Queloz2000,
Ohta2005,Winn2005,Hirano2011}. The spin-orbit angle distribution may be
a useful discriminator of the planetary migration scenarios.  Indeed the
RM effect has been measured successfully for around 70 transiting
exoplanets as illustrated in Figure \ref{fig:rmangle} below.  Among
them, 31 systems exhibit significant misalignment($\lambda>\pi/8$),
including 12 polar- and 7 retrograde-orbits \citep[see
also][]{Addison2013}.  These unexpected and counter-intuitive
discoveries imply that close-in giant planets should have experienced
violent dynamical processes, for instance, planet-planet scattering.

There is also a possibility that the spin-orbit
misalignment may be imprinted even at the initial condition of the
proto-planetary disk \citep[e.g.,][]{Lai2011,Batygin2013}, but a
reliable estimate of the distribution function of the initial spin-orbit
angles is challenging. Thus, 
in this paper, we focus on the dynamical process among the star and 
planet in the later stage of planet formation.

If the dynamical process is the major path to form hot Jupiters, one may expect that the spin-orbit angle distribution just after the formation of hot Jupiters should be very broad, and even close to random. In order to be consistent with the observed distribution with some overabundance around alignment configuration, a fairly efficient physical process of the spin-orbit (re)alignment is required. While the
subsequent tidal interaction between the central star and the innermost
planet is generally believed to be responsible for the alignment, its
detailed model is still unknown.

A conventional equilibrium tide model could realign the system, but
inevitably accompanies the orbital decay of the planet within a similar
timescale \citep{Barker2009, L2009, Matsumura2010}; see equation
(\ref{eq:tau-e}) below. Therefore a simple equilibrium tide model does
not explain the majority of the realigned systems with finite semi-major
axes.

In order to solve this problem, \citet{Winn2010} proposed a decoupling
model in which the stellar convective envelope is weakly coupled to its
radiative core, thus preferentially reducing the realignment timescale
of the stellar envelope relative to that of the orbital decay.  In
reality, however, it is not clear to what extent the decoupling model
can explain the observed distribution since there is no systematic and
quantitative study of this model.  In addition, the Sun is known to have
a strong coupling between the convective and radiative zones
\citep[e.g.,][]{Howe2009}, and theoretical support for this model seems
weak.

Recently, \citet{Lai2012} proposed a new model in which the damping
timescale of the spin-orbit angle could be significantly smaller than
that of the planetary orbit. When the stellar spin is misaligned with
respect to the planetary orbital axis, one component in the tidal
potential may excite inertial waves in the convective zone and provide a
dynamical tidal response \citep[e.g.,][]{Goodman2009}. This effect
increases the efficiency of the realignment process without contributing
to the orbital decay.

The model was studied in detail by \citet{Rogers2013}, who numerically
integrated a set of simplified equations for the semi-major axis $a$,
the spin angular frequency $\Omega_{\rm s}$, and the spin-orbit angle
$\Theta$, originally derived by \citet{Lai2012}. They found that
planetary systems with initially arbitrary spin-orbit angles have three
stable configurations; parallel, anti-parallel, and polar orbits.

In the present paper, we use a full set of equations to trace the
three-dimensional orbit of a planet (we assume that this represents the
innermost planet in the case of multi-planetary systems) and the spin
vector of the central star simultaneously, and examine the tidal
evolution of exoplanetary systems on a longer time-scale.  As a result,
we find that both anti-parallel and polar orbits eventually approach the
parallel orbit. We present a detailed comparison with the previous
results by \citet{Rogers2013}, and argue that the full set of equations
is important in understanding the longer-term evolution of the tidal
model.

We briefly review the two tidal models, the equilibrium tide and the
inertial wave dissipation, in \S \ref{sec:model}, and describe how to
combine the two models in our treatment. The basic set of equations is
summarized explicitly in \ref{sec:correia-eqs}. After comparing with the
previous result in \S \ref{subsec:comparison}, we present the evolution
in the combined tidal model in \S \ref{subsec:Lai-model}. We also
consider there the dependence on the fairly uncertain parameters of the
model.  Section \ref{sec:discussion} is devoted to summary and discusses
implications of the present result.

\section{Tidal evolution of star--hot Jupiter systems \label{sec:model}}

The entire dynamical evolution of planetary systems has many unresolved
aspects including the initial structure of proto-planetary disks,
formation processes of proto-planets, planetary migration, planet-planet
gravitational scattering, and the tidal interaction between the central
star and orbiting planets. It is definitely beyond the scope of the
present paper to consider such complicated processes in a
self-consistent fashion. Thus we consider a very simple system
comprising a star and a hot Jupiter, and focus on their tidal
interaction in order to examine the dynamical behavior of the stellar
spin and the planetary orbit. Just for definiteness, we fix the mass and
radius of the star and the planet as $m_{\rm s}=1M_\odot$, $m_{\rm
p}=10^{-3}M_\odot$, and $R_{\rm s}=1R_\odot$. The initial semi-major axis of the planetary orbit is set as
$a_{\rm ini}=0.02$AU.

 We do not assume any specific formation mechanism, but the above
configuration is expected generically from any successful models for hot
Jupiter formation. The eccentricity and inclination of the hot Jupiter
would depend on the details of the formation mechanism. The present
study, however, focuses on the tidal evolution between the star and the
hot Jupiter after the orbit circularization. Then we set the
eccentricity of the planet to zero, and consider a wide range of the
initial inclination of its orbit with respect to the spin axis of the
star. 

We numerically solve a set of equations \citep{Correia2011} for the stellar spin angular momentum $\mathbf S=I_{\rm s}{\mathbf \Omega}_{\rm s}$ ($I_{\rm s}$ is the
inertia moment of the star and ${\mathbf \Omega}_{\rm
s}$ is the spin angular frenquency), and the orbital
angular momentum $\mathbf L=m_{\rm p}m_{\rm s}/(m_{\rm
p}+m_{\rm s})a^2{\mathbf \Omega}_{\rm p}$ (${\mathbf
\Omega}_{\rm p}$ is the orbital mean motion) in order to explore the
tidal evolution of the system. \citet{Correia2011} derived those
equations for the equilibrium tide model (ET model, hereafter), which is
based on the weak-friction model with a constant delay time
$\Delta{t_{\rm L}}$ \citep{Mignard1979}.

We summarize the full equations in \ref{sec:correia-eqs} that directly
trace the three dimensional orbit of planets.  If one neglects the
eccentricity, the general relativity effect, and the tidal deformation
of the planet, however, they lead to the following simplified equations
for the semi-major axis $a$, the spin angular frequency $\Omega_{\rm
s}$, and the spin-orbit angle $\Theta$ in the ET model \citep{Lai2012}:
\begin{eqnarray} 
\label{eq:adot-e} 
\frac{(\dot a)_{\rm e}}{a} 
&=& -\frac{1}{\tau_{\rm e}}
\left(1-\frac{\Omega_{\rm s}}{\Omega_{\rm p}}\cos\Theta\right) ,\\
\label{eq:omegasdot-e} 
\frac{(\dot\Omega_{\rm s})_{\rm e}}{\Omega_{\rm s}} 
&=& \frac{1}{\tau_{\rm e}}
\left(\frac{L}{2S}\right)
\left[\cos\Theta-\frac{\Omega_{\rm s}}{2\Omega_{\rm p}}
(1+\cos^{2}\Theta)\right] ,\\
\label{eq:thetadot-e} 
(\dot\Theta)_{\rm e}
&=& -\frac{1}{\tau_{\rm e}}
\left(\frac{L}{2S}\right)
\sin\Theta \left[1-\frac{\Omega_{\rm s}}{2\Omega_{\rm p}}
\left(\cos\Theta-\frac{S}{L}\right)\right] .
\end{eqnarray}
The subscript e refers to the term due to the ET model.

As is clear from equations (\ref{eq:adot-e}) to (\ref{eq:thetadot-e}),
$a$, $\Omega_{\rm s}$, and $\Theta$ have similar damping timescales
characterized by
\begin{eqnarray} 
\label{eq:tau-e}
\tau_{\rm e} &=& 
\left(\frac{Q_{\rm e}}{3k_{2}}\right)
\left(\frac{m_{\rm s}}{m_{\rm p}}\right)
\left(\frac{a}{R_{\rm s}}\right)^5 \Omega_{\rm p}^{-1} \cr
&\approx& 1.3 
\left(\frac{Q_{\rm e}/3k_{2}}{2\times10^6}\right)
\left(\frac{m_{\rm s}}{10^3m_{\rm p}}\right)
\left(\frac{\bar\rho_{\rm s}}{\bar\rho_\odot}\right)^{5/3}
\left(\frac{P_{\rm p}}{1 {\rm day}}\right)^{13/3} {\rm Gyr},
\end{eqnarray}
where $\bar\rho_{\rm s}$ is the mean density of the star, $P_{\rm p}$ is
the orbital period of the planet, $Q_{\rm e} \equiv (2\Omega_{\rm p}
\Delta t_{\rm L})^{-1}$ is the tidal quality factor of the star, and
$k_{2}$ is the Love number of the star that represents gravity field
deformation at the stellar surface in response to an external perturbing
potential of spherical harmonic degree 2. We note that $k_2$ depends
only on the internal density distribution of the star. The specific
values of $\Delta{t_{\rm L}}$ and $k_{2}$ that we employ are shown in Table \ref{tab:simparam}.

The fact that the ET model is basically governed by the single timescale
is inconsistent with the presence of a number of well-aligned hot
Jupiters orbiting at a finite distance from the star, as long as the ET
model is the dominant mechanism for the spin-orbit (re)alignment.  This
is why \citet{Winn2010} proposed a decoupling model in which the star's
convective envelope is weakly coupled to its radiative core and $\Theta$
is damped efficiently without significant orbital decay of the planet.

Another model that we consider in detail below is proposed by
\citet{Lai2012}, who pointed out the importance of the inertial waves of
the star that are driven by the Coriolis force and dissipated by the
tidal interaction of the misaligned $\mathbf S$ and $\mathbf L$.  More
specifically, he expanded the tidal potential due to the planet orbiting
at $\mathbf r$ in the spherical coordinate system $(r, \theta, \phi)$
centered at the star with the $z$-axis along the stellar spin $\mathbf
S$:
\begin{eqnarray}
\label{eq:def-U2}
U_2(\mathbf{r},t)
= - \sum_{m, m'}U_{m m'}(m_{\rm p}, \Theta)r^2
Y_{2m}(\theta,\phi)e^{-im'\Omega_{\rm p}t} ,
\end{eqnarray}
where the index $m'$ refers to that of spherical harmonics
$Y_{2m'}(\theta_{\rm L}, \phi_{\rm L})$ defined in the coordinate with
its $z$-axis along the planetary orbital angular momentum
$\mathbf{L}$. Then he found that the only additional
tidal torque in this model comes from $(m, m')=(1,0)$ component.

The corresponding tidal torque components due to such
dynamical tides are given in \citet{Lai2012} as 
\begin{eqnarray} 
\label{eq:Tx10} 
T_{10,x} &=& 
\frac{3k_{10}}{4Q_{10}}T_{0}\sin{\Theta}\cos^{3}{\Theta}, \\
\label{eq:Ty10} 
T_{10,y} &=&
-\frac{3k_{10}}{16}T_{0}\sin{4\Theta},\\ 
\label{eq:Tz10} 
T_{10,z} &=& 
-\frac{3k_{10}}{4Q_{10}}T_{0}(\sin{\Theta}\cos{\Theta})^{2}, 
\end{eqnarray}
where $k_{10}$ and $Q_{10}$ are the dimensionless tidal Love number and
tidal quality factor corresponding to the (1,0) component, and
\begin{eqnarray} 
T_{0}= G \left(\frac{m_{\rm p}}{a^3}\right)^2 R_{\rm s}^5 
\end{eqnarray}
\citep[c.f.,][]{Murray1999}.

We note here that \citet{Lai2012} chooses the $y$-direction along the
direction ${\mathbf S} \times {\mathbf L}$. Thus $T_{10,y}$ does not
change the spin-orbit elements, contributing to the spin precession, but
we compute $T_{10,y}$ on the basis of equation (20) of \citet{Lai2012} in
any case. Up to the leading-order of the delay time, $T_{10,y}$ does not
depend on $Q_{10}$ unlike $T_{10,x}$ nor $T_{10,z}$.

From the corresponding tidal torques due to the inertial
wave dissipation described above, \citet{Lai2012} derived the following
equations for $a$, $\Omega_{\rm s}$, and $\Theta$:
\begin{eqnarray} 
\label{eq:adot-10}
(\dot a)_{10}&=&0, \\
\label{eq:omegasdot-10}
\frac{(\dot\Omega_{\rm s})_{10}}{\Omega_{\rm s}}
&=& -\frac{1}{\tau_{10}} \left(\sin\Theta\cos\Theta\right)^{2}, \\
\label{eq:thetadot-10} 
(\dot\Theta)_{10}
&=& -\frac{1}{\tau_{10}}\sin\Theta\cos^{2}\Theta
\left(\cos\Theta+\frac{S}{L}\right) .
\end{eqnarray}
In the above equations, $\tau_{10}$ is the characteristic tidal damping
timescale corresponding to the (1,0) component\footnote{We adopt the
definition of $\tau_{10}$ by \citet{Rogers2013}, which corresponds to
$t_{\rm s10}$ of \citet{Lai2012}.}:
\begin{eqnarray} 
\label{eq:tau-10}
\tau_{10} 
&=& \left(\frac{4Q_{10}}{3k_{10}}\right)
\left(\frac{m_{\rm s}}{m_{\rm p}}\right)
\left(\frac{a}{R_{\rm s}}\right)^{5}
\left(\frac{S}{L}\right)\Omega_{\rm p}^{-1} \cr
&=& 4 \left(\frac{Q_{10}}{Q_{\rm e}}\right)
\left(\frac{k_{2}}{k_{10}}\right)
\left(\frac{S}{L}\right)
\tau_{\rm e} .
\end{eqnarray}

Thus the inertial wave dissipation adds the damping terms with the
timescale of $\tau_{10}$ for $\Theta$ and $\Omega_s$, while the
planetary orbit is unaffected since the (1,0) component of the tidal
potential is static in the inertial frame \citep{Lai2012}.  Thus in this
model the spin-orbit angle aligns faster before the planet falls into
the central star, or more strictly, closer to its Roche limit $\approx
2R_{\rm s}(\bar\rho_{\rm s}/\bar\rho_{\rm p})^{1/3}$, if $\tau_{10}$ is
much smaller than $\tau_{\rm e}$. The value of $\tau_{10}/\tau_{\rm e}$,
however, is difficult to estimate in a reliable fashion. Thus we assume
a fiducial value of $10^{-3}$ at the start of our simulation following
\citet{Rogers2013}, and examine the dependence in \S
\ref{subsec:Lai-model}.

We numerically integrate the set of equations shown in
 \ref{sec:correia-eqs} for the ET model. For the Lai model, we
 modify the equations by adding the tidal torque $\mathbf T_{10}$ as
\begin{eqnarray}
\label{eq:L0}
\dot{\mathbf S} &=& \dot{\mathbf S}_{\rm (e)} + {\mathbf T}_{10} 
- {\mathbf T}_{10,\rm e},\\
\label{eq:G1}
\dot{\mathbf L} &=& \dot{\mathbf L}_{\rm (e)} - {\mathbf T}_{10} 
+ {\mathbf T}_{10,\rm e},
\end{eqnarray}
where the subscript (e) indicates the terms for the ET model, and
${\mathbf T}_{10,\rm e}$ is the term that is introduced to avoid the
double counting in the above equations:
\begin{eqnarray}
\label{eq:DoubleCounting}
\frac{{\mathbf T}_{10,\rm e}}{{\mathbf T}_{10}} 
&=& \left(\frac{Q_{10}}{Q_{\rm e}}\right)
\left(\frac{k_{2}}{k_{10}}\right).
\end{eqnarray}

Table \ref{tab:simparam} summarizes the fiducial values
that we employ in the simulations.  We use the same values for the
stellar principal moment of inertia, the Love number and the tidal delay
time that were adopted by \citet{Correia2011} for $\sim 1M_\odot$
star.

\begin{table}[t]
\begin{center}
\begin{tabular}{c|c|c|c} 
Parameter &symbol& Star & hot Jupiter\\    \hline
mass&$m[M_{\odot}]$&1 & $10^{-3}$\\
radius &$R[R_{\odot}]$& 1 & --- \\
principal moment of inertia&$C/mR^{2}$   & 0.08 &---\\ 
Love number &$k_{2}$  & 0.028 & --- \\
tidal delay time&$\Delta{t_{\rm L}}[\rm sec]$ & 0.1 & ---\end{tabular}
\caption{Fiducial values of the parameters in our simulations.}
\label{tab:simparam}
\end{center}
\end{table}

\section{Numerical results \label{sec:result}}

\subsection{Comparison with previous results 
\label{subsec:comparison}}

Before presenting a detailed analysis of the Lai model, we compare
typical results arising from different tidal models and approximations.
We consider two different tidal models: the ET model and the Lai
model. Unlike in \citet{Rogers2013}, we refer to the Lai model which
incorporates both the equilibrium tide and the inertial wave dissipation
effects.

We numerically integrate the full set of equations in
\ref{sec:correia-eqs} throughout the paper.  If one focuses on the evolution for $a$,
$\Omega_{\rm s}$, and $\Theta$, the evolution equations in the Lai model 
can be reduced as follows:
\begin{eqnarray}
\label{eq:adot-lai}
\dot a &=& (\dot a)_{\rm e}, \\
\label{eq:omegasdot-lai} 
\dot\Omega_{\rm s} &=& (\dot\Omega_{\rm s})_{\rm e}
+  (\dot\Omega_{\rm s})_{10} - (\dot\Omega_{\rm s})_{10,\rm e},\\
\label{eq:thetadot-lai} 
 \dot\Theta &=& (\dot\Theta)_{\rm e} 
  + (\dot\Theta)_{10} - (\dot\Theta)_{10,\rm e} ,
\end{eqnarray}
where 
\begin{eqnarray}
\label{eq:DoubleCounting2}
\frac{(\dot\Omega_{\rm s})_{10,\rm e}}{(\dot\Omega_{\rm s})_{10}}  =\frac{(\dot\Theta)_{10,\rm e}}{(\dot\Theta)_{10}} = \left(\frac{Q_{10}}{Q_{\rm e}}\right)
\left(\frac{k_{2}}{k_{10}}\right).
\end{eqnarray}
In order to compare with the results from the full set of equations, we also integrate the above set of equations assuming the constant value
for $(Q_{10}/Q_{\rm e})(k_{2}/k_{10})$. In practice, we use
equation (\ref{eq:tau-10}), and fix its value from the initial values of
$\tau_{10}/\tau_{\rm e}$ and $S/L$:
\begin{eqnarray}
\label{eq:qk}
  \frac{Q_{10}}{Q_{\rm e}}\frac{k_{2}}{k_{10}}=
\frac{1}{4}\left(\frac{\tau_{10}}{\tau_{\rm e}}\right)_{\rm ini} 
\left(\frac{S}{L}\right)^{-1}_{\rm ini} .
\end{eqnarray}
In the above simplified set of equations, $m_{\rm s}$ and $m_{\rm p}$ do
not show up explicitly, but implicitly in $\tau_{10}$, $\tau_{\rm e}$, and
$I_{\rm s}$. For definiteness, we fix $I_s=0.08m_{\rm s}R_{\rm s}^2$ \citep{Wu2003, Correia2011}.
Thus $S/L$ and $\Omega_{\rm s}/\Omega_{\rm p}$ are related to each other
as
\begin{eqnarray}
\label{eq:omegasp}
\frac{\Omega_{\rm s}}{\Omega_{\rm p}}
\approx 0.23
\left(\frac{M_\odot}{m_{\rm s}}\right)
 \left(\frac{m_{\rm p}}{10^{-3}M_\odot}\right)
\left(\frac{R_\odot}{R_{\rm s}}\right)^2
\left(\frac{a}{0.02{\rm AU}}\right)^2
\left(\frac{S}{L}\right) .
\end{eqnarray}

\begin{figure}[t]
\begin{center}
\includegraphics[width=12cm]{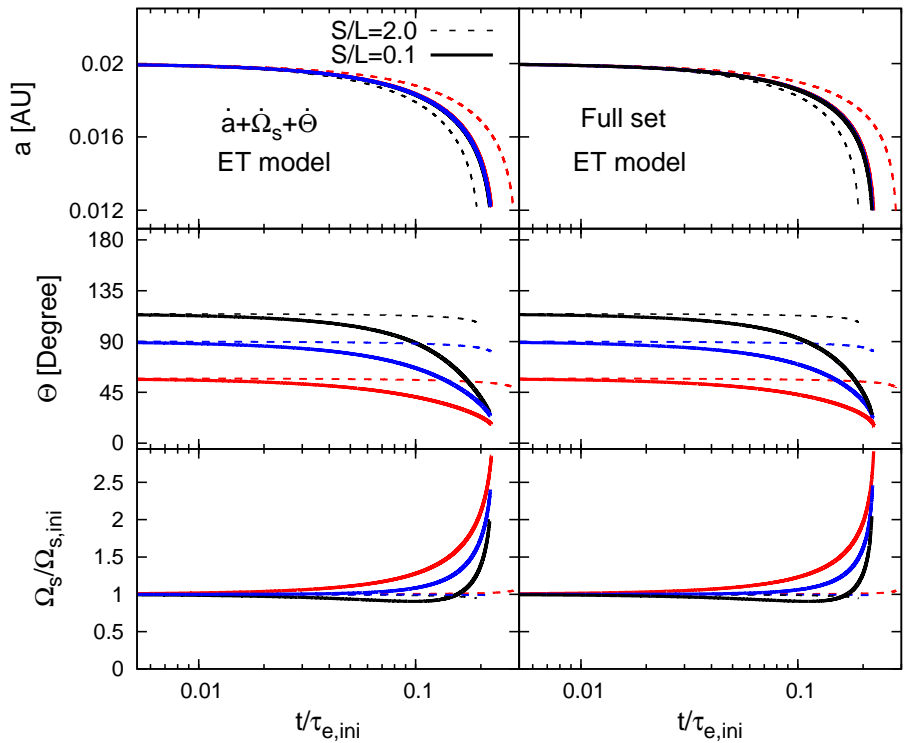} 
\caption{Evolution of the semi-major axis ({\it Top}), the spin-orbit
angle ({\it Center}), and the spin angular frequency ({\it Bottom}) in
the equilibrium tide model.  Left and right panels are based on the
numerical integration of the simplified and full sets of equations
explained in the text.  The three different initial spin-orbit angles
are assumed; prograde ($\Theta_{\rm ini}=58^{\circ}$), polar
($\Theta_{\rm ini}=90^{\circ}$), and retrograde ($\Theta_{\rm
ini}=116^{\circ}$) orbits, shown in red, blue, and black, respectively.
The solid and dashed lines indicate the $(S/L)_{\rm ini}=0.1$ (solid) and $2$ (dashed).} \label{fig:etmodel}
\end{center}
\end{figure}

Figure \ref{fig:etmodel} shows evolution of $a$ ({\it Upper} panels),
$\Theta$ ({\it Center}), and $\Omega_{\rm s}/\Omega_{\rm s,ini}$ ({\it
Lower}) for the ET model.  Left and right panels plot the results on the
basis of the simplified set of equations for $a$, $\Omega_{\rm s}$ and
$\Theta$ (eqs.(\ref{eq:adot-e}) -- (\ref{eq:thetadot-e})) and the full
equations in \ref{sec:correia-eqs}, respectively.  Note that $\tau_{\rm
e}$ completely specifies the units of time in the simulations. Thus the
time evolution is scaled with respect to the initial value of $\tau_{\rm
e}$. This is why those panels are plotted against $t/\tau_{\rm e, ini}$.

In each panel, we start the models with three different initial
spin-orbit angles; prograde ($\Theta_{\rm ini}=58^{\circ}$), polar
($\Theta_{\rm ini}=90^{\circ}$), and retrograde ($\Theta_{\rm
ini}=116^{\circ}$).  They are shown in red, blue, and black lines,
respectively.  For each initial condition, we plot two cases with the
$(S/L)_{\rm ini}=0.1$ (solid) and $2$ (dashed) roughly corresponding to
the upper and lower limits of $S/L$ for planets observed via the RM
effect \citep{Rogers2013}.  We note here that the system evolved towards
$\Theta=0$ regardless of $\Theta_{\rm ini}$.

 As pointed out earlier by several authors, the spin-orbit alignment
occurs almost simultaneously with the planetary orbital decay.  More
strictly, equations (\ref{eq:adot-e}) and (\ref{eq:thetadot-e}) indicate
that the damping time-scales of the orbit and the spin-orbit angle are
given by $\tau_{\rm e}$ and $(S/L)\tau_{\rm e}$, respectively.  This
explains the dependence on $(S/L)_{\rm ini}$ in the plots; compare solid
and dashed lines.

\begin{figure}[t]
\begin{center}
\includegraphics[width=12cm]{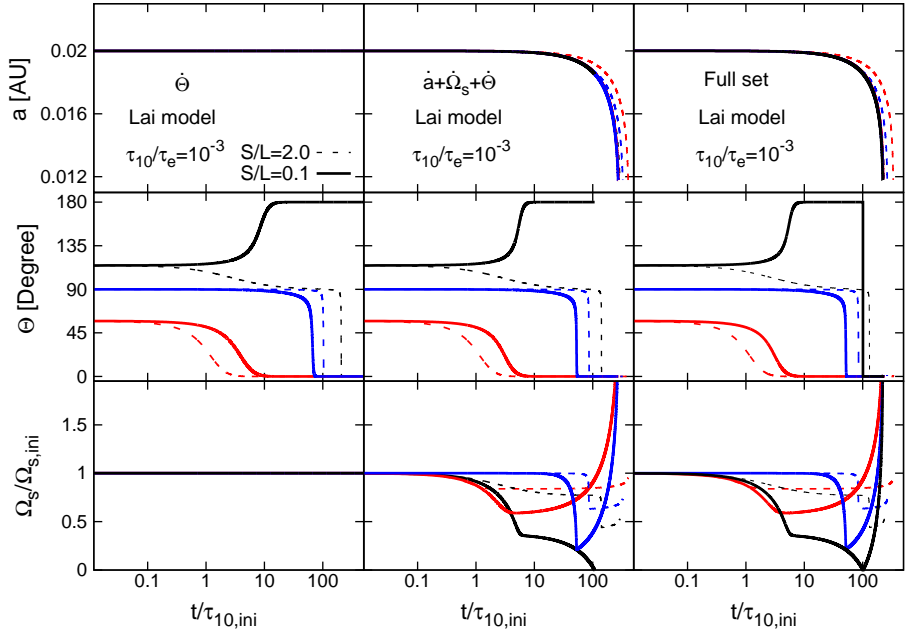} 
\caption{Same as Figure
\ref{fig:etmodel} but for the Lai model that combines the equilibrium
tide and the inertial wave dissipation effects simultaneously.  We add
left panels that correspond to the result neglecting the change of $a$
 and $\Omega_{\rm s}$.} \label{fig:laimodel}
\end{center}
\end{figure}

The situation changes drastically in the Lai model, which introduces an
additional time-scale $\tau_{10}$. Figure \ref{fig:laimodel} shows an
example of $(\tau_{10}/\tau_{\rm e})_{\rm ini}=10^{-3}$. In this case,
the orbital decay proceeds according to the time-scale of $\tau_{\rm
e}$, while the alignment time-scale is controlled by $\tau_{10}$ as well
as $(S/L)\tau_{\rm e}$. Thus if $\tau_{10} \ll \tau_{\rm e}$, one can
neglect the change of $a$ during the spin-orbit evolution. This is the
approximation adopted by \citet{Rogers2013}, which corresponds to the
left panels of Figure \ref{fig:laimodel}; we numerically integrate
equation (\ref{eq:thetadot-lai}) neglecting the time evolution of $a$
and $\Omega_s$.

\citet{Rogers2013} found that the system has three distinct stable
configurations, {\it i.e.}, anti-parallel ($\Theta=\pi$), polar ($\pi/2$),
and parallel ($0$) orbits, which is easily expected from the
right-hand-side of equation (\ref{eq:thetadot-lai}) as
well. Nevertheless the damping on the order of $\tau_{\rm e}$ shows up
in the later stage, and the polar orbit evolves towards $\Theta=0$.
This can be hardly recognized in Figure 2 of \citet{Rogers2013} because
we suspect they stop the integration at $t=100\tau_{10}=0.1\tau_{\rm
e}$, before their approximation $\dot a=0$ becomes invalid.

In reality, the evolution beyond the epoch should be computed taking
account of the change of $a$ and $\Omega_{\rm s}$ properly. The
middle panels show the result, and confirm that the polar orbit
is a metastable configuration.  Since the right-hand-sides of both
equations (\ref{eq:thetadot-e}) and (\ref{eq:thetadot-lai}) are
proportional to $\sin\Theta$, the system should evolve eventually
towards either $\Theta=0$ or $\pi$, but not $\pi/2$.

We found that the $\Theta=\pi$ configuration finally approaches
$\Theta=0$ by integrating the full set of equations for the three
dimensional planetary orbit.  In the anti-parallel case, the spin-orbit
angle has a sharp change between $\pi$ and $0$.  This is because as the
orbit damps, the stellar spin $S$ continuously decreases according to
the total angular momentum conservation.  At some point, therefore, the
stellar spin becomes 0, changes the direction, and then starts to
increase (aligned to the orbital axis). Thus the really stable
configuration is $\Theta=0$ alone. Nevertheless the
duration of such meta-stable configurations is also sensitive to the
choice of $\tau_{10}/\tau_{\rm e}$ and/or $(S/L)_{\rm ini}$; see Figure
\ref{fig:tau-dependence}. Thus the retro-grade and polar-orbit systems
can be observed in the real systems depending on their age.

This behavior cannot be traced properly by the simplified
approach, which is based on the differential equation
for $\Theta$, combining equations (\ref{eq:thetadot-e}) and
(\ref{eq:thetadot-10}); the right-hand-side of equations (\ref{eq:thetadot-e})
diverges at $S=0$ or $\Omega_s=0$, and cannot be numerically integrated beyond the point. In contrast, the full set of equations in
\ref{sec:correia-eqs} computes ${\mathbf S}$ and ${\mathbf L}$ first, and $\Theta$ later. Thus one can compute the evolution continuously
beyond ${\mathbf S}=0$. This is one of the advantages of using the full
set of equations even in the case of the simple star-planet system.

In any case, we confirm the original claim of \citet{Lai2012} that
one can have an aligned system with a finite semi-major axis as long as
$\tau_{10}/\tau_{\rm e} \ll 1$ is satisfied.

\subsection{Spin-orbit angle evolution in the Lai model 
\label{subsec:Lai-model}}

\begin{figure}[t]
\begin{center}
\includegraphics[width=12cm]{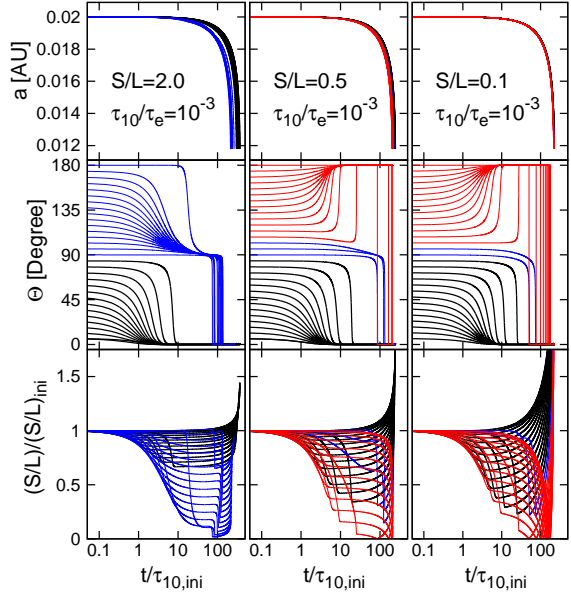} 
\caption{Evolution of the semi-major axis ({\it Top}), the spin-orbit
angle ({\it Middle}), and $S/L$ ({\it Bottom}) in the Lai model for 30
systems with $\Theta_{\rm ini}$ from $0^{\circ}$ to $180^{\circ}$ with
interval $6^{\circ}$.  {\it Left}, {\it center}, and {\it right} panels
indicate $(S/L)_{\rm ini}=2$, $0.5$, and $0.1$, respectively. The
$(\tau_{10}/\tau_{\rm e})_{\rm ini}=10^{-3}$ is assumed.  Blue and red
lines indicate the systems that show the transition from polar to
parallel, anti-parallel to parallel states, respectively.}
\label{fig:full-Lai}
\end{center}
\end{figure}

Now we examine the Lai model more systematically using the full set of
equations. We run 30 models of a planet with a regularly spaced
$\Theta_{\rm ini}$ between $0^{\circ}$ and $180^{\circ}$.  We plot the
results in Figure \ref{fig:full-Lai} for $(S/L)_{\rm ini}=2$ (left),
$0.5$ (middle), and $0.1$ (right). All the simulations adopt
$(\tau_{10}/\tau_{\rm e})_{\rm ini}=10^{-3}$ so as to compare the middle
panels of Figure 2 of \cite{Rogers2013}.

As explained in \S \ref{subsec:comparison}, the system first approaches
parallel, polar, or anti-parallel orbits within a time-scale of
$\tau_{10}$. They are plotted in black, blue, and red, respectively, so
that they are easily distinguished visually.  Next the polar, and
subsequently anti-parallel orbits, approach towards the parallel orbits
in a time-scale of $\tau_{\rm e}$, eventually falling below the Roche
limit of the star (0.012 AU in the present case), where we stop the
simulations.

The transition from $\Theta=180^\circ$ to $0^\circ$ (red curves) happens
through a state of $\Omega_{\rm s}=0$, implying that the stellar spin
starts counter-rotating due to the tidal effect of the orbiting
planet. As mentioned in \S \ref{subsec:comparison}, the evolution beyond
$\Omega_{\rm s}=0$ is difficult to trace with the simplified set of
equations. Thus our simulations on the basis of the full set of
equations are essential.  Note that the transition to the three
meta-stable configurations is fairly rapid.  Therefore if the age of the
system is larger than $\tau_{10}$ and smaller than $\tau_{\rm e}$, one
may expect basically three distinct spin-orbit angles, but their broad
distribution as observed ({\it c.f.,} Figure \ref{fig:rmangle} below) is
not likely to be explained even taking into account the
projection effect;see Figure 3 of \citet{Rogers2013}.

While Figure \ref{fig:full-Lai} is the main result of the paper, it
remains to consider the dependence on the initial ratios of
$(\tau_{10}/\tau_{\rm e})_{\rm ini}$, and $(\Omega_{\rm s}/\Omega_{\rm
p})_{\rm ini}$. As we will show below, the behavior presented in Figure
\ref{fig:full-Lai} is indeed robust against the choice of those
parameters.

\begin{figure}[t]
\begin{center}
\includegraphics[width=12cm]{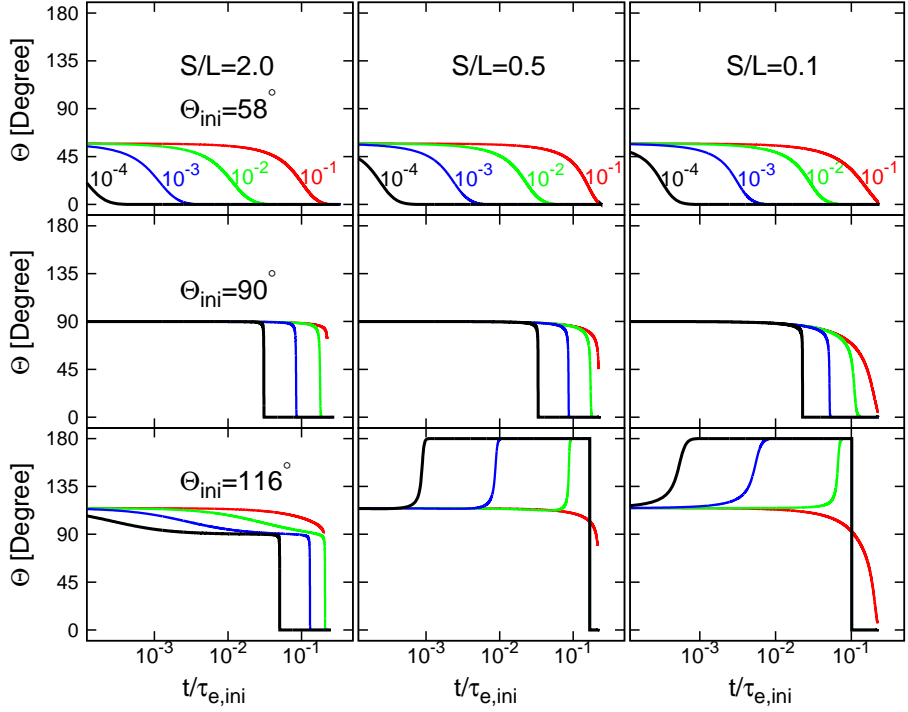}
\caption{Dependence on the tidal dissipation parameter
 $(\tau_{10}/\tau_{\rm e})_{\rm ini}$ for $(S/L)_{\rm ini}=2.0$ ({\it
 Left}), $0.5$ ({\it Middle}), $0.1$ ({\it Right}). Black, Blue, Green,
 Red line represent $(\tau_{10}/\tau_{\rm e})_{\rm ini} = 10^{-4},
 10^{-3}, 10^{-2}, 10^{-1}$; {\it Top}, {\it center}, and {\it bottom}
 panels indicate initially prograde, polar, and retrograde orbits,
 respectively.  }\label{fig:tau-dependence}
\end{center}
\end{figure}

Figure \ref{fig:tau-dependence} shows results of initially prograde
({\it Top}), polar ({\it Center}), and retrograde ({\it Bottom}), orbits
with $(S/L)_{\rm ini}=2.0$, $0.5$, and $0.1$ for $(\tau_{10}/\tau_{\rm
e})_{\rm ini}=10^{-4}$ (black), $10^{-3}$ (blue), $10^{-2}$ (green), and
$10^{-1}$ (red).  The ratio $\tau_{10}/\tau_{\rm e}$ reflects the
property of the stellar fluid itself, and thus is not easy to predict in
a reliable fashion. Therefore we consider a fairly wide range of its
possible value.  The initially prograde cases approach $\Theta=0$ with a
time-scale of $\tau_{10}$ fairly clearly.  The initially polar-orbit
cases stay the configuration for a significantly longer period than
$\tau_{10}$, but eventually approach $\Theta=0$. The initially
retrograde cases are somewhat intermediate between the two.  In any
case, the behavior changes systematically with the value of
$(\tau_{10}/\tau_{\rm e})_{\rm ini}$ and can be
interpolated/extrapolated relatively easily from Figure
\ref{fig:tau-dependence}.

\begin{figure}[t]
\begin{center}
\includegraphics[width=14cm]{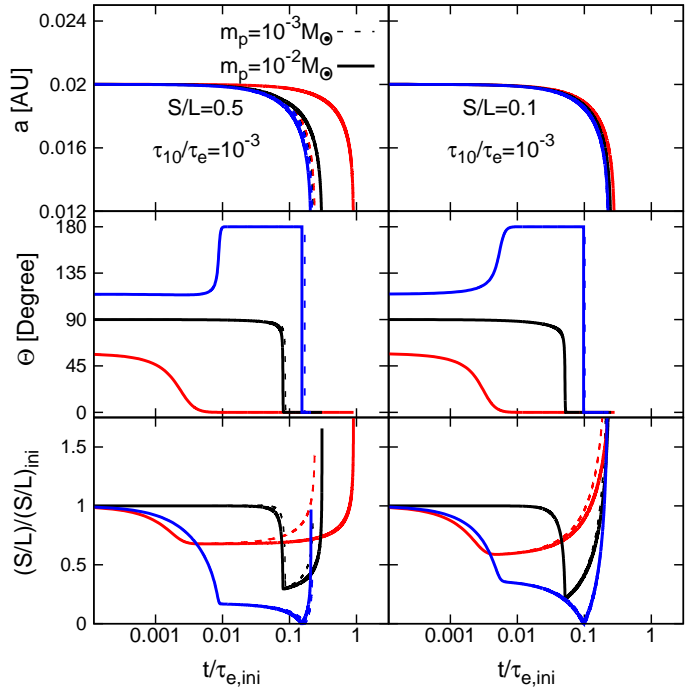}
\caption{Dependence on the planet mass $m_{\rm p}$, or
equivalently on $(\Omega_{\rm s}/\Omega_{\rm p})_{\rm ini}$, for
$(S/L)_{\rm ini}=0.5$ ({\it Left}), $0.1$ ({\it Right}). The blue,
black and red lines correspond to initial retrograde, polar and prograde
orbit, while the solid and dashed lines to $m_{\rm p} = 10^{-2}M_\odot$
and $10^{-3}M_\odot$, respectively.}\label{fig:w-dependence}
\end{center}
\end{figure}

Figure \ref{fig:w-dependence} shows the dependence on the planet mass
$m_{\rm p}$, or equivalently on $(\Omega_{\rm s}/\Omega_{\rm p})_{\rm
ini}$ through equation (\ref{eq:omegasp}).  Since our simulations adopt
$m_{\rm s}=1M_\odot$, $R_{\rm s}=1R_\odot$, and $m_{\rm
p}=10^{-3}M_{\odot}$, equation (\ref{eq:omegasp}) determines the value
of $\Omega_{\rm s}/\Omega_{\rm p}$ uniquely through $(S/L)_{\rm
ini}$. In contrast, \citet{Rogers2013} vary the value randomly in the
range of $0.1 <(\Omega_{\rm s}/\Omega_{\rm p})_{\rm ini}<10$, while they
do not describe exactly how.  Equation (\ref{eq:omegasp}) implies that
the corresponding values of $(\Omega_{\rm s}/\Omega_{\rm p})_{\rm ini}$
in our model with $m_{\rm p}=10^{-3}M_\odot$ are $0.46$, $0.11$, and
$0.02$ for $(S/L)_{\rm ini}=2$, $0.5$, and $0.1$, respectively.  In
order to see the dependence, we simply repeat the simulations with
$m_{\rm p}=10^{-2}M_{\odot}$, keeping the other parameters exactly the
same. Thus the $m_{\rm p}=10^{-2}M_{\odot}$ case corresponds to an order
of magnitude increase of $(\Omega_{\rm s}/\Omega_{\rm p})_{\rm ini}$
relative to the $m_{\rm p}=10^{-3}M_{\odot}$ case.  We
do not run the case with $m_{\rm p}=10^{-2}M_{\odot}$ and $(S/L)_{\rm
ini}=2$ since it does not satisfy the criterion $\Omega_{\rm p}\gg
\Omega_{\rm s}$, under which the $(m,m')=(1,0)$ is the only excitation mode
\citep{Lai2012}.  Figure \ref{fig:w-dependence} basically shows that the
result depends on $(\Omega_{\rm s}/\Omega_{\rm p})_{\rm ini}$ very
weakly.

\section{Summary and discussion \label{sec:discussion}}

We have considered tidal evolution of star--hot Jupiter systems with
particular attention to their spin-orbit alignment. We focused on the
inertial wave dissipation model proposed by \citet{Lai2012}, and
examined the extent to which the model reproduces the observed
distribution of spin-orbit angles for transiting exoplanets.

Basically we confirmed the conclusion of \citet{Rogers2013} that the Lai
model has three distinct stable configurations, anti-parallel, polar,
and parallel orbits. In reality, however, the former two turn out to be
meta-stable, and approach the parallel orbits over a longer time-scale
as the equilibrium tide effect exceeds that of the inertial wave
dissipation. We also found that the later evolution stage needs to be
examined using direct three dimensional integrations for the spin
vector and orbital angular momentum vector of the
system, rather than the simplified differential equations for the
semi-major axis, spin angular frequency, and spin-orbit angle, even when
a simple star--planet system is considered.

The relative importance of the two tidal effects is determined by the
ratio of $\tau_{\rm e}$ and $\tau_{10}$. Unfortunately there is a huge
uncertainty in predicting the value of each parameter. Nevertheless in
order to achieve a spin-orbit alignment at a finite planetary orbit,
$\tau_{\rm 10} \ll \tau_{\rm e}$ is required. In such cases, however,
the alignment due to the inertial wave dissipation works too
efficiently, and the observed broad distribution of $\Theta$ for
transiting planets (more precisely, the projected angle $\lambda$ of
$\Theta$ onto the sky plane) is difficult to reproduce.

\begin{figure}[t]
\begin{center}
\includegraphics[width=12cm]{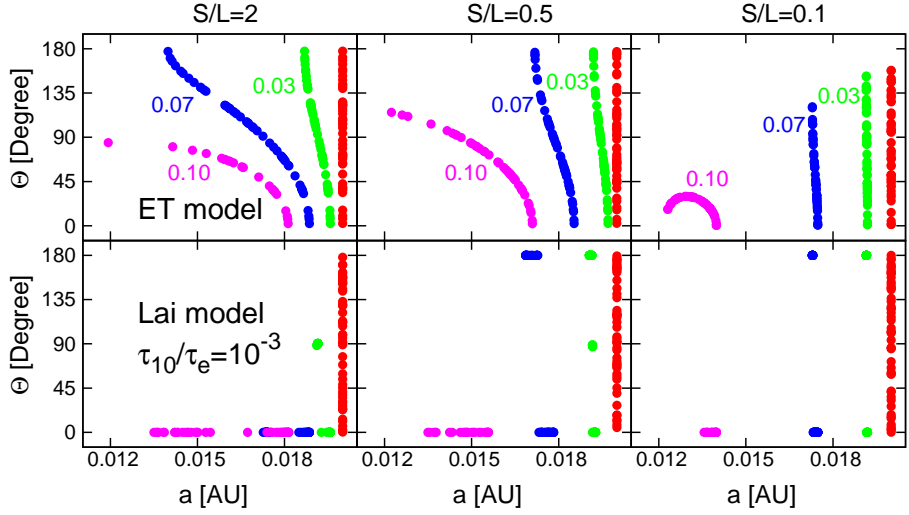}
\caption{Spin-orbit angle plotted against the semi-major axis at
different epochs; $t=0$ (red), $t=0.03\tau_{\rm e}$ (green),
$t=0.07\tau_{\rm e}$ (blue), and $t=0.1\tau_{\rm e}$ (magenta).  The
{\it upper} and {\it lower} panels correspond to the equilibrium tide
and Lai models, respectively, for 50 systems with randomly distributed
$\Theta_{\rm ini}$. The initial values of $S/L$ are $2$ ({\it Left}),
$0.5$ ({\it Center}), and $0.1$ ({\it
Right}).}\label{fig:correlation-a-theta}
\end{center}
\end{figure}

To illustrate this point, we simulate 50 systems with a planet located
initially at 0.02AU and a randomly chosen $\Theta_{\rm ini}$ for
$(S/L)_{\rm ini}=2$, $0.5$, and $0.1$.  Figure
\ref{fig:correlation-a-theta} plots resulting $\Theta$ against $a$ at
four different epochs; $t/\tau_{\rm e,ini}=0$, $0.03$, $0.07$, and
$0.1$.  While the ET model ({\it Upper}) predicts a relatively
continuous correlation between $\Theta$ and $a$, the Lai model ({\it
Lower}) has the distinct three (meta-)stable states, but they
subsequently become completely aligned by $t=0.1\tau_{\rm e}$. The
evolution is so rapid that even the different value of the initial
semi-major axis cannot broaden the distribution significantly.

\begin{figure}[t]
\begin{center}
\includegraphics[width=12cm]{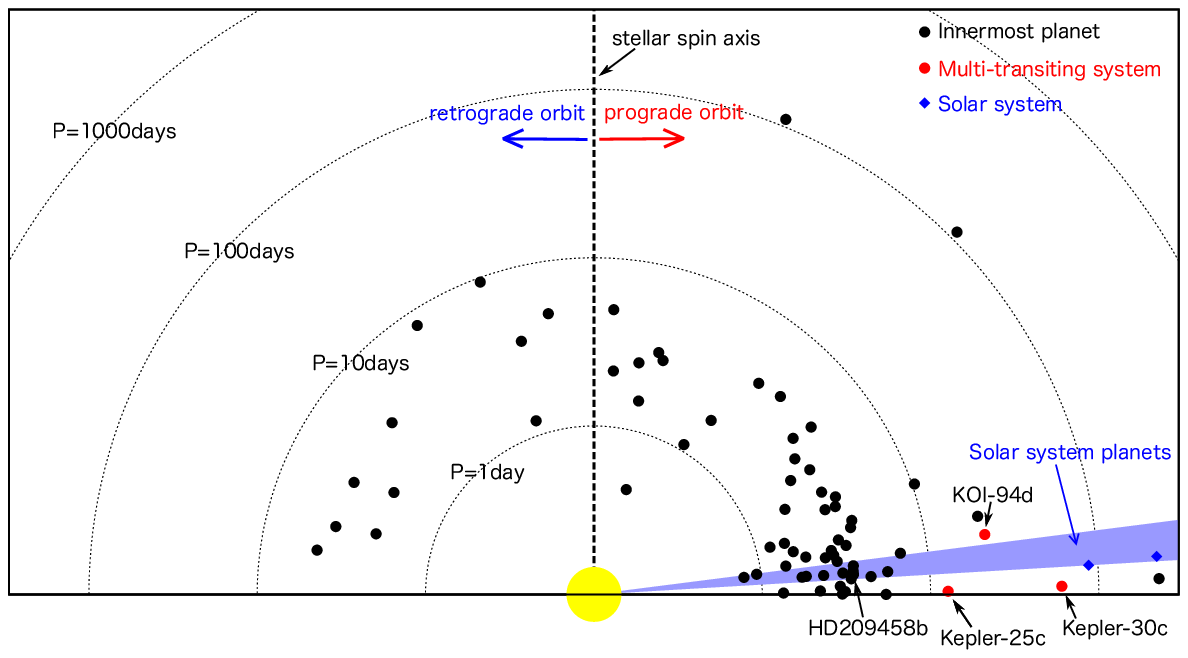} \caption{The projected
misalignment angles $\lambda$ for transiting planets as of August 20, 2013.  Black circles
indicate innermost planets in single transiting systems, while red
circles denote the largest planets in the three multi-transiting
systems, Kepler-25c\citep{Albrecht2013}, Kepler-30c\citep{SO2012}, and
KOI-94d\citep{Hirano2012,Masuda2013}.}\label{fig:rmangle}
\end{center}
\end{figure}

The observed distribution of the projected spin-orbit angles $\lambda$
is plotted in Figure \ref{fig:rmangle} on the basis of the
Holt-Rossiter-McLaughlin Encyclopedia compiled by Ren\'e
Heller\footnote{http://www.physics.mcmaster.ca/{\textasciitilde}rheller/};
the radial coordinate of each symbol corresponds to the logarithm of the
orbital period of the planet ($P \propto a^{3/2}$), and its angular
coordinate represents the observed value of $\lambda$.  Black and red
circles indicate the innermost planets in single transiting systems, and
the largest planets in the multi-transiting systems, respectively. The
range of the solar-system planets is plotted in a blue region.

There is a clear tendency of clumping around $0<\lambda<30^\circ$, in
addition to the dominance of the prograde orbits relative to the
retrograde ones. Nevertheless the distribution is rather broad, and does
not seem to be consistent with that expected from Figure
\ref{fig:correlation-a-theta}.  In order to make more quantitative
comparison, one needs to consider the effect of projection since
$\lambda$ is different from $\Theta$, but corresponds to its projected
angle on the sky. The effect, however, does not change the main
conclusion, and we will leave the detailed analysis to our future study.

In this sense, we agree with the overall conclusion of
\citet{Rogers2013} that the Lai model, at least in the current simple
star--planet system, cannot explain the wide range of observed
$\lambda$.  Furthermore, the current model is unlikely to explain the
empirical trend that the realigned systems are preferentially found in
the host stars with $T_{\rm eff} < 6250$ K
\citep{Winn2010}.  Nevertheless we have to recognize that the planetary
system considered in this paper is oversimplified; we ignore the
outer planets that may influence the dynamics of the innermost planet
significantly and the host-star dependence of the tidal parameters. In
addition, we totally neglect the dependence on the initial conditions before the tidal realignment. Thus it is premature to make a negative conclusion, and we
should explore the wider range of system configurations. For this
purpose, the simplified set of equations is inappropriate, and we need
to integrate three dimensional orbits of multi-planets using the full
set of equations \citep{Correia2011}. Also it is important and
interesting to examine three dimensional evolution of the spin and
orbital angular momenta directly, instead of that of their mutual angle
alone.

\acknowledgments

We thank Shoya Kamiaka for his careful reading of the manuscript,
and a referee for very valuable and pertinent comments,
which improved the earlier manuscript.  Y.S. gratefully acknowledges
the supports from the Global Collaborative Research Fund¡ÈA Worldwide
Investigation of Other Worlds¡Égrant, the Global Scholars Program of
Princeton University, and the Grant-in Aid for Scientific Research by
JSPS (No. 24340035).  K.M. is supported by the Leading Graduate Course
for Frontiers of Mathematical Sciences and Physics.  T.H. is supported
by Japan Society for Promotion of Science (JSPS) Fellowship for Research
(No. 25-3183).  A.T. acknowledges the support from Grant-in-Aid for
Scientific Research by JSPS (No. 24540257).

\clearpage
\appendix
\renewcommand\thesection{\appendixname~\Alph{section}}
\renewcommand\theequation{\Alph{section}.\arabic{equation}}

\section{Basic equations for tidal evolution
\label{sec:correia-eqs}}

The present paper considers a hybrid tidal model which combines a
conventional equilibrium tide model with the inertial tidal dissipation
model proposed by \citet{Lai2012}. Just for completeness, we explicitly
give a full set of basic equations that we numerically integrate.  The
equations are derived by \citet{Correia2011} for the equilibrium tide
model for a system of a central star and two planets with arbitrary
orbital eccentricities. The tidal interaction between the stellar spin
and the planetary orbit is based on the weak-friction model with a
constant delay time $\Delta{t}$ (Mignard 1979).  Furthermore the general
relativity correction and the stellar oblateness are taken into account.

In the present paper, we focus on the tidal interaction
between the star and the innermost planet. Thus we neglect the distant
planet for simplicity. The tide on the planet is not considered either
because it should have a very negligible effect on the dynamics of the
star and planet. Furthermore since we fix the initial eccentricity as 0
in the current simulations, the GR effect is also neglected.

Let us denote their mass by $m_{\rm s}$ and $m_{\rm p}$, the semi-major axis by $a$, the spin angular velocity by $\Omega_{\rm s}$, the
orbital angular velocity by $\Omega_{\rm p}$, radius by $R_{\rm s}$ and gravity
coefficient by $J_{2_{s}}$.  The equations are written in terms of the
Jacobi coordinates with $\mathbf r_{1}$ being the relative position from
$m_{\rm s}$ to $m_{\rm p}$ under the quadrupole approximation for the
conservative and tidal potentials of the system
\citep[e.g.][]{Smart1953,Kaula1964,Correia2011}.

The evolution of spin of star and orbit of planet can be
specified in the quadrupole approximation by two parameters;

1) star rotational angular momentum:
\begin{equation} 
\label{eq:Li} 
\mathbf S=C_{\rm s}\Omega_{\rm s}\hat{\mathbf s},
\end{equation}
where $\hat{\mathbf s}$ is the unit vector of $\hat{\mathbf S}$,
and $C_{\rm s}$ is the principal moment of inertia.

2)orbital angular momentum:
\begin{equation} 
\label{eq:Gi} 
\mathbf L=\beta \sqrt{{\mu} a}\hat{\mathbf k},
\end{equation}
where $\hat{\mathbf k}$ is the unit vector of $\hat{\mathbf L}$, $\mu=G(m_{\rm s}+m_{\rm p})$, and $\beta=m_{\rm s}m_{\rm p}/(m_{\rm s}+m_{\rm p}$).

We define $\Theta$, the angle between the spin of the star and the orbit of the innermost planet via
\begin{equation} 
\label{eq:thetai} 
\cos\Theta=\hat{\mathbf s}\cdot \hat{\mathbf k} .
\end{equation}

As \citet{Correia2011} described in detail, the evolution of the system
is governed by the conservative motion and the tidal effects. First, the equation for conservative
motion is obtained by averaging the equations of motion over the mean
anomaly:
\begin{eqnarray} 
\label{eq:G1dotc} 
\dot{\mathbf L}_{(\rm c)}
&=& \alpha\cos\Theta
\hat{\mathbf s}\times\hat{\mathbf k}, \\
\label{eq:Lidotc} 
\dot{\mathbf S}_{(\rm c)} &=& 
-\alpha\cos\Theta \hat{\mathbf s}\times\hat{\mathbf k},
\end{eqnarray}
where $L$ is the norm of $\mathbf L$, and
\begin{eqnarray} 
\label{eq:alpha1i} 
\alpha&=&\frac{3Gm_{\rm s}m_{\rm p}J_{2_{s}}R_{s}^{2}}
{2a^{3}} .
 \end{eqnarray}

Second, the equilibrium tidal effect is considered under the quadrupole
approximation of the tidal potential assuming the constant delay time
$\Delta{t_{\rm L}}$ \citep{Mignard1979}.  After averaging
equations of motion over the mean anomaly, one obtains
\begin{eqnarray} 
\label{eq:Gdott} 
 \dot{\mathbf L}_{(\rm t)} &=&
-\dot{\mathbf S}_{(\rm t)}, \cr
\dot{\mathbf S}_{(\rm t)}
&=& K  
\Big[ \frac{\Omega_{\rm s}}{2}
(\hat{\mathbf s}- \cos\Theta\hat{\mathbf k})
 - \Omega_{\rm s} \hat{\mathbf s}
 + \Omega_{\rm p} \hat{\mathbf k}\Big] ,
\end{eqnarray}
where

\begin{eqnarray} 
\label{eq:Ki}   
K&=&\Delta{t_{\rm L}}
 \frac{3k_{2}Gm_{\rm p}^{2}R_{\rm s}^{5}}{a^{6}},
\end{eqnarray}

Thus the total rates of change of $L$ and $S$ under
the equilibrium tidal model are given by the sum of the above
terms corresponding to the conservative motion and tidal effect:
\begin{eqnarray} 
\label{eq:Gie} 
\dot{\mathbf L}_{(\rm e)} 
&=& \dot{\mathbf L}_{(\rm c)} +\dot{\mathbf L}_{(\rm t)}, \\
\label{eq:Lie}
\dot{\mathbf S}_{(\rm e)} 
&=&\dot{\mathbf S}_{(\rm c)} +\dot{\mathbf S}_{(\rm t)}.
\end{eqnarray}

Finally, we add the tidal torque due to the inertial wave dissipation,
equations (\ref{eq:Tx10}) to (\ref{eq:Tz10}) to the above equations as
in equations (\ref{eq:L0}) and (\ref{eq:G1}). 

\clearpage


\end{document}